%% file: main.tex
\documentclass[acmsmall,screen]{acmart}

\usepackage{hyperref}
\usepackage[breakable]{tcolorbox}
\usepackage{listings}
\usepackage{xspace}
\usepackage{algorithm}
\usepackage{algpseudocode}
\usepackage{makecell}
\usepackage{enumitem}
\usepackage{graphicx}
\usepackage{booktabs}
\usepackage{xparse}
\usepackage{soul}
\usepackage{multirow}
\usepackage{tikz}

\usetikzlibrary{fit,calc}

\makeatletter
\long\def\@mpfootnotetext#1{%
  \global\setbox\@mpfootins\vbox{%
    \unvbox\@mpfootins
    \reset@font\footnotesize
    \hsize\columnwidth
    \@parboxrestore
    \protected@edef\@currentlabel
         {\csname p@mpfootnote\endcsname\@thefnmark}%
    \color@begingroup%\centering - commented out centering footnotes
      \@makefntext{%
        \rule\z@\footnotesep\ignorespaces#1\@finalstrut\strutbox}%
    \color@endgroup}}
\makeatother

\setlist[itemize]{leftmargin=1.5em}
\setlist[enumerate]{leftmargin=2em}
\input{macro}

\setcopyright{cc}
\setcctype{by}
\acmDOI{10.1145/3797084}
\acmYear{2026}
\acmJournal{PACMSE}
\acmVolume{3}
\acmNumber{FSE}
\acmArticle{FSE056}
\acmMonth{7}
\received{2025-09-12}
\received[accepted]{2025-12-22}

\begin{document}

%%
%% The "title" command has an optional parameter,
%% allowing the author to define a "short title" to be used in page headers.
\title{Reducing Cost of LLM Agents with Trajectory Reduction}

\author{Yuan-An Xiao}
\orcid{0000-0002-5673-3831}
\affiliation{%
  \department{Key Lab of HCST, MOE, School of Computer Science}
  \institution{Peking University}
  \city{Beijing}
  \country{China}
}
\email{xiaoyuanan@pku.edu.cn}

\author{Pengfei Gao}
\orcid{0000-0003-3800-2565}
\affiliation{%
  \institution{ByteDance}
  \city{Beijing}
  \country{China}
}
\email{gaopf1995@gmail.com}

\author{Chao Peng}
\orcid{0000-0003-2843-0689}
\affiliation{%
  \institution{ByteDance}
  \city{Beijing}
  \country{China}
}
\email{pengchao.x@bytedance.com}

\author{Yingfei Xiong}
\authornote{Corresponding author.}
\orcid{0000-0001-8991-747X}
\affiliation{%
  \department{Key Lab of HCST, MOE, School of Computer Science}
  \institution{Peking University}
  \city{Beijing}
  \country{China}
}
\email{xiongyf@pku.edu.cn}

%%
%% By default, the full list of authors will be used in the page
%% headers. Often, this list is too long, and will overlap
%% other information printed in the page headers. This command allows
%% the author to define a more concise list
%% of authors' names for this purpose.
%\renewcommand{\shortauthors}{Trovato et al.}

\input{sections/0_abstract}

%\received{20 February 2007}
%\received[revised]{12 March 2009}
%\received[accepted]{5 June 2009}

% BEGIN space-saving during review
%\setcopyright{none}
%\settopmatter{printacmref=false}
% END space-saving

%%
%% This command processes the author and affiliation and title
%% information and builds the first part of the formatted document.
\maketitle

\input{sections/1_intro}
\input{sections/2_motivation}
\input{sections/3_approach}
\input{sections/4_eval}
\input{sections/5_discussion}
\input{sections/6_related}
\input{sections/7_ending}

%%
%% The acknowledgments section is defined using the "acks" environment
%% (and NOT an unnumbered section). This ensures the proper
%% identification of the section in the article metadata, and the
%% consistent spelling of the heading.

%%
%% The next two lines define the bibliography style to be used, and
%% the bibliography file.
\bibliographystyle{ACM-Reference-Format}
\bibliography{bib/zotero,bib/ref}

\end{document}

%% file: macro.tex
\makeatletter
\def\code{%
  \begingroup
  \small\ttfamily
  \@activecode\@code
}
\def\@activecode{%
  \catcode`.\active \catcode`_\active \catcode`'\active \catcode``\active
}
\begingroup\@activecode
\gdef\@code#1{%
  \def.{\char46\allowbreak}%
  \def_{\_\allowbreak}%
  \let'\textquotesingle
  \let`\textasciigrave
  #1%
  \endgroup
}
\endgroup
\makeatother

\algnewcommand\algorithmicforeach{\textbf{for each}}
\algdef{S}[FOR]{ForEach}[1]{\algorithmicforeach\ #1\ \algorithmicdo}

\newtcolorbox{rqbox}{breakable,leftrule=1mm,rightrule=1mm,toprule=0mm,bottomrule=0mm,left=6pt,right=6pt,top=6pt,bottom=6pt,before upper=,colback={orange!8},colframe={orange!80}}
\newtcolorbox{infobox}[1]{breakable,left=1pt,right=1pt,top=2pt,bottom=2pt,title={#1},before upper=}

\newcommand{\ourtool}{AgentDiet\xspace}

\newcommand{\figref}[1]{Fig.~\ref{fig:#1}}
\newcommand{\tabref}[1]{Table~\ref{tab:#1}}
\newcommand{\secref}[1]{Section ~\ref{sec:#1}}
\newcommand{\algoref}[1]{Algorithm~\ref{alg:#1}}

\newcommand{\chapquote}[3]{\vspace{.25em} \begin{quotation} \textit{#1} \end{quotation} \begin{flushright} - #2, \textit{#3}\end{flushright} \vspace{1em}}

%% file: sections/0_abstract.tex
\begin{abstract}
Multi-turn agent systems based on Large Language Models (LLMs) have become increasingly popular for software engineering tasks. While LLM agents demonstrate promising effectiveness, the high computational cost of input tokens due to ever-growing trajectories remains a significant efficiency concern. Efficiency has been largely overlooked in existing studies and agent products, and this paper addresses this gap by introducing an inference-time trajectory reduction approach that reduces computational costs.

By analyzing existing agent trajectories, we demonstrate that useless, redundant, and expired information is widespread across trajectories. Such waste can be identified and reduced without compromising the agent's performance. We propose a simple yet effective trajectory reduction approach, \ourtool, which automatically removes such waste during agent execution. We implement \ourtool on a top-performing coding agent, and our evaluation on two LLMs and two benchmarks shows that \ourtool can reduce input tokens by 39.9\%--59.7\% and the total computational cost by 21.1\%--35.9\%, while maintaining the same agent performance. These results indicate that inference-time trajectory reduction is a promising direction for agent systems.
\end{abstract}


\ccsdesc[500]{Software and its engineering~Software evolution}
\ccsdesc[500]{Software and its engineering~Software testing and debugging}

\keywords{LLM Agent, Code Language Model, Trajectory Reduction}

%% file: sections/1_intro.tex
\section{Introduction}

Large Language Models (LLMs) have been widely used for various software engineering tasks such as code generation, testing, and repair~\cite{liu2024large,jiang2023impact,farzandway2025automated,ahmed2024automatic,xia2024agentless,xia2024automated}. A prominent trend in the application of LLMs is agent-based approaches~\cite{bouzenia2024repairagent,qin2024agentfl,kang2025explainable,wang2024openhands,yang2024sweagent,zhang2024autocoderover}, where LLMs are asked to complete complex tasks through multi-step reasoning and tool calling. For example, mini-SWE-agent~\cite{yang2024sweagent} achieves a single-shot performance of 65\% in SWE-bench Verified~\cite{jimenez2024swebench} using Claude 4 Sonnet~\cite{claude4}, indicating that agents can fix real-world GitHub issues with basic tools such as file editing and bash scripting. Given their promising performance, LLM agents are integrated into various AI products~\cite{claude_code,cursor}, and 24\% of professional developers already use LLM agents daily or weekly in 2025~\cite{stackoverflow_survey}.

Although these agent approaches demonstrate promising performance in terms of effectiveness, efficiency remains a concern for their application~\cite{fan2025swe}. In a recent survey by StackOverflow~\cite{stackoverflow_survey}, 53\% of the participants believe that the cost of using AI agents is a barrier for them. This efficiency issue has been largely overlooked in existing studies, leaving a significant gap to address.

The root cause of this efficiency issue lies in the workflow of LLM agents, specifically the ever-growing trajectory. In a typical agent system, once the LLM calls a tool, the tool call and its result are kept in the trajectory until the agent completes the entire task. Therefore, if the LLM opens a large file or runs a command that generates verbose output, the computational cost of these tokens snowballs in each subsequent step~\cite{swe_effi}. This results in non-negligible waste. For example, in September 2025, the daily token usage of Claude 4 Sonnet reached 100 billion on the OpenRouter platform~\cite{openrouter_claude4_usage}, and 99\% of them were input tokens accumulated in the trajectory, while only the remaining 1\% were generated by the LLM.

To address the above efficiency issue, this paper focuses on reducing trajectory tokens for coding LLM agents. In academia, most existing papers on token reduction focus on single-turn tasks, such as trimming the retrieved input for question-answering~\cite{pan2024llmlingua2,chirkova2025provence,wang2023learning,louis2025pisco,zhang2022diet}. These tasks differ from agentic coding in multiple aspects, limiting the applicability of existing approaches. Specifically: (1) these approaches reduce all input tokens at once, whereas trajectory tokens gradually accumulate in agents, making the timing of reduction a new research question; (2) these approaches reduce tokens in natural language, while coding agents process structured information such as software source code; (3) some approaches require modifying the LLM inference process or even the model itself, which is impractical and unsuitable for proprietary LLMs that do not provide such capabilities. To the best of our knowledge, there is no existing publication on inference-time trajectory reduction for coding agents. In industry, agent products such as Claude Code and Cursor~\cite{claude_code,cursor} sparingly apply LLM-based trajectory reduction (often called ``compaction'') only when the context window is full or the user explicitly requests it, focusing on improving robustness rather than efficiency. As a result, the potential for efficiency improvement via trajectory reduction remains unknown. This paper reveals this potential by designing an LLM-based trajectory reduction approach and evaluating it on multiple benchmarks and LLMs.

In \secref{motivation}, we analyze the problem and investigate the potential for cost reduction, revealing that the trajectories of LLM agents contain a significant amount of waste, including useless, redundant, or expired information. Based on this empirical finding, we propose a prototype approach to reduce this waste using an LLM. In \secref{approach}, we instantiate this design into a concrete algorithm, \ourtool. \secref{impl} discusses the integration of \ourtool into agents and the effect of hyperparameter settings through a quantitative experiment. \secref{evaluation} evaluates \ourtool on two LLMs and two benchmarks, showing that it can steadily reduce input tokens by 39.9\%--59.7\%, or the final computational cost by 21.1\%--35.9\%, while maintaining the same agent performance. \secref{discussion} discusses possible future work and threats to the validity of the evaluation. \secref{related_work} reviews related work.

The contributions of this paper are summarized as follows:

\begin{itemize}
    \item We reveal that inference-time trajectory reduction of coding agents is a promising direction, with evaluation results showing that significant cost reduction is possible without harming agent performance through frequent trajectory reduction during agent execution;
    \item We propose \ourtool, a simple yet effective trajectory reduction approach, which is open-source and can be easily integrated into coding agents;
    \item We discuss the design of possible trajectory reduction approaches through case studies and large-scale quantitative experiments.
\end{itemize}

\begin{rqbox}
The artifacts of \ourtool are available at \href{https://doi.org/10.6084/m9.figshare.30073654}{https://doi.org/10.6084/m9.figshare.30073654}.
\end{rqbox}

%% file: sections/2_motivation.tex
\section{Problem Analysis}
\label{sec:motivation}

In \secref{motivating-preliminary}, we first discuss preliminary concepts related to this paper, i.e., the typical workflow of LLM agents and the contents of their trajectories. We motivate our approach in \secref{motivation-waste-examples} and \secref{motivation-design} by empirically identifying typical waste in trajectories and then designing a prototype trajectory reduction approach to address it.

In particular, we aim to answer a series of questions:

\begin{itemize}
    \item Is there waste in the trajectories? \emph{(Yes, there is a lot.) $\to$ \secref{motivation-waste-examples}}
    \item Can we automatically identify and reduce the waste in trajectories? \emph{(Yes, it is possible to use an LLM for trajectory reduction.) $\to$ \secref{motivation-design-reduce}}
    \item Do LLMs know when to perform trajectory reduction on their own? \emph{(No, so we enforce the trajectory reduction step with a separate reflection module.) $\to$ \secref{motivation-design-timing}}
    \item Is calling an LLM for this purpose ideal and efficient? \emph{(Not really, but the overhead is under control with simple optimizations.) $\to$ \secref{motivation-design-overhead}}
\end{itemize}

\subsection{Preliminaries}
\label{sec:motivating-preliminary}

\chapquote{``Agents are models using tools in a loop.''}{Hannah Moran from Anthropic}{Prompting for Agents}

\begin{figure}
    \centering
    \includegraphics[width=0.7\linewidth]{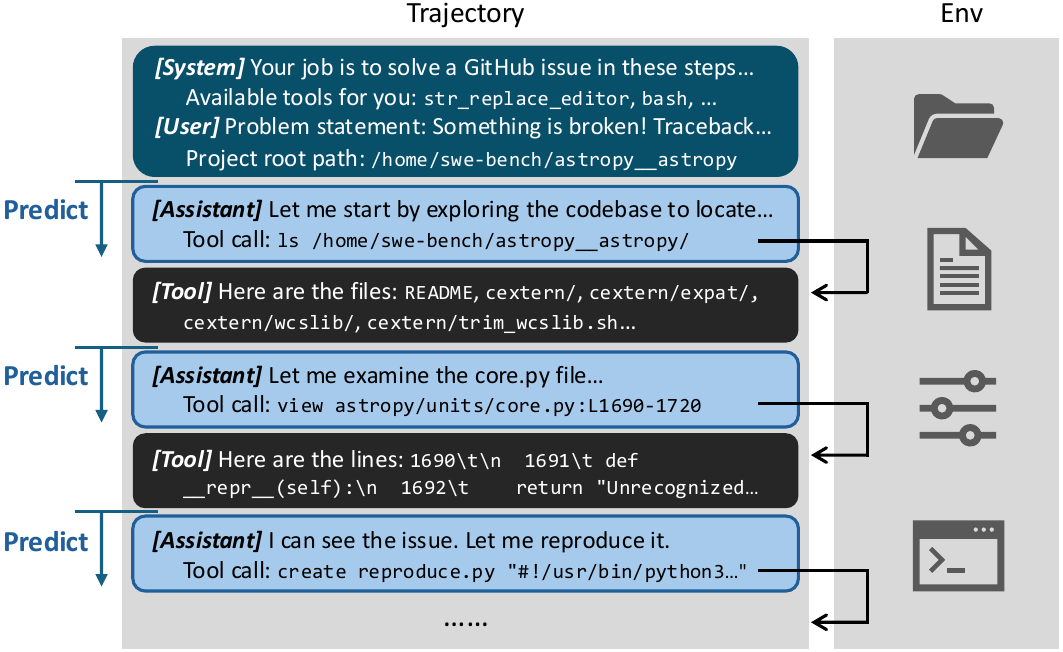}
    \caption{Typical workflow of an LLM agent.}
    \label{fig:motivation-traj}
\end{figure}

In recent years, Large Language Models (LLMs) have demonstrated the ability to assist with a wide variety of tasks. Following established approaches such as Chain-of-Thought~\cite{wei2022chain}, ReAct~\cite{yao2023react}, and CodeAct~\cite{wang2024executable}, LLM agents have emerged as a promising paradigm for integrating LLMs into such tasks by equipping the LLM with tools to interact with the environment. In the field of software engineering, this means that the LLM can operate on the codebase autonomously with the ability to read files, make edits, execute commands, commit changes, etc.

\figref{motivation-traj} shows the typical workflow of an LLM agent. The core of an agent is its \emph{trajectory}, which records all the necessary context that will be given to the LLM. An agent starts with an initial trajectory that contains the instruction to the LLM (in a system message) and the current task (in a user message). The LLM then analyzes the current situation and predicts the next action to take in a standard tool call format (in an assistant message). The agent system parses the generated tool call, executes the action, and returns the output to the LLM (in a tool message). In this loop, assistant and tool messages are continuously concatenated into the trajectory, which will be the input to subsequent LLM predictions. The loop continues until the LLM decides to finish the task with a special indicator, or when the agent system reaches the step limit.

From this workflow, we can identify an obvious efficiency issue in such agent systems: Once the LLM calls a tool, the corresponding assistant and tool messages will be concatenated into the trajectory and will be kept forever until the task finishes. A long message is included in each subsequent LLM prediction, even if its content is irrelevant (or no longer relevant) to the task, leading to inefficient use of computational resources. As a quantitative result, the average trajectory for solving a single GitHub issue that we collected from the SWE-bench Verified~\cite{jimenez2024swebench} benchmark contains 48.4K tokens in 40 steps. Breaking down these tokens, tool messages (containing the results of tool calls) use 30.4K tokens, assistant messages use 13.7K tokens (of which 11.9K are the arguments of tool calls), and system/user messages (containing the initial instructions) use 4.4K tokens. Since each token concatenated into the trajectory is included in every subsequent input to the LLM, the accumulated token usage per issue reaches 1.0M, which requires excessive computational and memory resources.

Recent LLMs are equipped with the KV Cache mechanism, which mitigates the issue of high computational cost by caching repeated calculations of the Key and Value matrices in the Transformer~\cite{vaswani2017attention} architecture. However, this does not eliminate the need for trajectory reduction because: (1) KV Cache only caches part of the computations, and the remaining calculations are still costly as the trajectory grows rapidly; (2) KV Cache consumes hardware resources like VRAM and I/O bandwidth, so reducing the trajectory length conserves these valuable resources. In fact, the existence of KV Cache makes designing trajectory reduction approaches more challenging, since modifying a token in the trajectory invalidates the cache for all subsequent tokens.

Long messages in the context window not only harm the efficiency but also cause the performance of LLMs to degrade~\cite{liu2023lost,li2025longcodeu}. Therefore, removing content from trajectories does not necessarily lead to a drop in the agent's performance. Instead, it has the potential to maintain or even slightly improve the performance of agents by removing waste.

\subsection{Waste in Trajectories}
\label{sec:motivation-waste-examples}

This subsection presents a pilot study where we analyze the feasibility of trajectory reduction by first identifying typical waste in trajectories. For this purpose, we use SWE-bench Verified~\cite{jimenez2024swebench}, a popular benchmark requiring agent systems to solve GitHub issues in well-known Python repositories. The benchmark makes the logs and trajectories of all participants available for download. We downloaded and manually inspected 100 trajectories from Trae Agent~\cite{gao2025trae}, which is based on Claude 4 Sonnet and ranked top on SWE-bench Verified.

By qualitatively inspecting the contents of the trajectories, we found that waste is widespread in almost all trajectories. We categorized the waste into three typical scenarios in which tokens could be removed or compressed from a human perspective:

\subsubsection{Useless Information}

Some information is irrelevant to the task and can be safely removed with minimal information loss. For example, nearly all trajectories begin with a tool call to enumerate all files in the repository, and the tool response includes cache and resource files (e.g., files under the \code{__pycache__} and \code{.egg-info} directories). In all trajectories, the agent also executed commands to build and test the project, which may generate verbose output (e.g., GNU \code{make} prints the message ``\code{make[2]: Entering/Leaving directory '...'}'' for each visited location by default). Although Trae Agent has a mechanism to truncate output to the first 16KB, a fixed threshold only avoids extreme cases, and we still frequently observed useless information within this threshold.

\subsubsection{Redundant Information}

When a piece of information appears multiple times in the trajectory, redundant copies can be removed without losing information. The most typical case of redundant information involves the \code{str_replace_editor} tool, a standard file editing tool designed by Anthropic that is frequently used in all trajectories. The tool call arguments for \code{str_replace_editor} are passed as JSON, such as \code{\{"command": "str_replace", "path": "F", "old_str": "P", "new_str": "Q"\}}, where P is a unique fragment of file F that typically contains multiple lines, and will be replaced by Q. The tool will perform the string replacement and then respond with the replaced result. There are multiple redundancies in such a tool call: (1) Q in the argument is repeated in the replaced result in the tool response; (2) Q may contain the same components (e.g., shared statements) as P; (3) the agent must have retrieved the context to edit prior to this tool call, so P repeats code from previous steps in the trajectory.

\subsubsection{Expired Information}

Information relevant to a local step may no longer be necessary after the step is completed. A typical case is where the agent loops through many items to find the relevant one. For example, in the process of diagnosing the root cause of the issue, the agent often searches for a symbol with the \code{grep} shell command, and then opens each file to read as it appears in the search results. This process takes a few steps in the trajectory, and once the agent identifies the faulty file, most content in other files becomes unnecessary.

\subsection{A Prototype Approach to Trajectory Reduction}
\label{sec:motivation-design}

\subsubsection{Reducing the Waste}
\label{sec:motivation-design-reduce}

In this subsection, we discuss how to automatically identify and reduce waste in trajectories. A straightforward approach would be to maintain a set of rules (such as regular expressions) for each typical kind of waste, but such rules are unlikely to cover the variety of scenarios. For example, it is difficult to write a rule to identify all verbose output in test logs, since every project has its own output format.

Given the heuristic nature of this task, we chose to identify and reduce waste with the help of an LLM. We designed a prompt for trajectory reduction consisting of four parts:

\begin{infobox}{Prompt outline for \ourtool}
\textbf{[1. High-level job description]}

You will analyze and compress a given step in the trajectory of an LLM agent solving a software bug. Your job is to \ldots{}

\smallskip

\textbf{[2. Input/output format]}

In the trajectory, each step is wrapped in <step id="..."></step>. You should process \ldots

\smallskip

\textbf{[3. Summary and examples of the three kinds of typical waste]}

Typical examples:

\hspace{.2em}- (Useless information) If the step opens a huge file but only one part is necessary \ldots

\hspace{.2em}- (Redundant information) If the step uses str\_replace\_editor to modify a file \ldots

\hspace{.2em}- \ldots

\smallskip

\textbf{[4. Guidelines to prevent information loss after reduction]}

\hspace{.2em}- Replace the text to remove with a short takeaway, e.g., \ldots

\hspace{.2em}- You should keep the original structure unchanged, e.g., XML tags, \ldots

\hspace{.2em}- \ldots
\end{infobox}

\begin{figure}
    \centering
    \includegraphics[width=.95\linewidth]{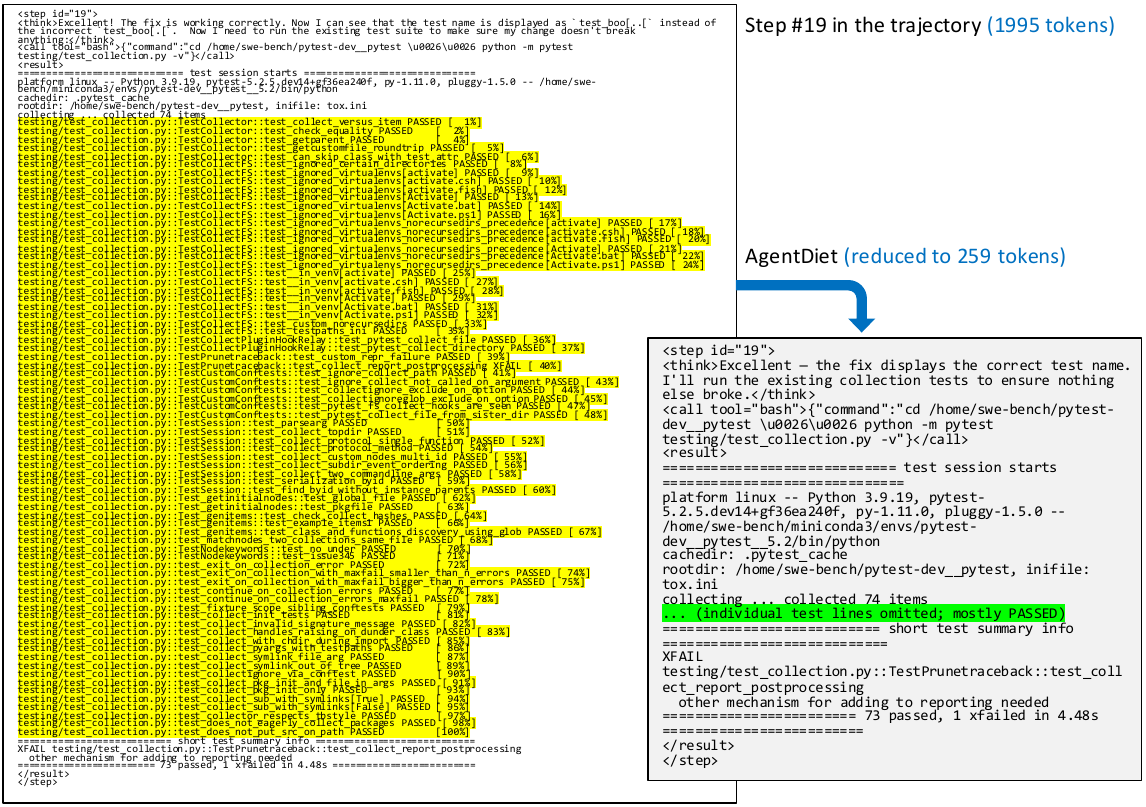}
    \caption{A case study of reducing information waste in the trajectory (pytest-dev\_\_pytest-6202).}
    \label{fig:motivation-casestudy}
\end{figure}

The examples in the third part came directly from the identified waste in the pilot study. The guidelines in the fourth part were gradually refined using the identified waste examples as test cases: Whenever we observed that the LLM failed to remove waste or deleted too much information, we refined the guidelines in the prompt until the LLM passed the test case. The final prompt is available in the artifact, and we omit it here due to space constraints.

\figref{motivation-casestudy} illustrates how this approach guides the LLM to reduce waste in the trajectory. It shows step 19 in a trajectory solving the SWE-bench instance ``\code{pytest-dev__pytest-6202}'' as an example. Here, the agent runs the test suite and retrieves its output, which contains the full list of tests (highlighted in yellow) followed by the only failing test. This example falls into the ``useless information'' category, because the verbose list of passed tests is unnecessary for the current task. We serialize this step into the XML format, where the waste costs 1714 tokens out of 1995 tokens. Based on our prompt, the LLM replaces the waste with a short placeholder (``individual test lines omitted; mostly PASSED'', highlighted in green) while preserving the important failing test. We can then update the agent's trajectory with the replaced version that has only 259 tokens.

\subsubsection{Agent Integration}
\label{sec:motivation-design-timing}

After showing that trajectory reduction is possible with an LLM, the next natural topic is how to integrate this into an agent system.

Ideally, if the agent system is powerful enough, it could reduce its own trajectory given a tool for that purpose. We tested this hypothesis by implementing a tool called \code{erase}, which allows the agent to override existing steps in the trajectory. We assigned a numeric ID to each segment in the trajectory, and the LLM could call the \code{erase} tool with arguments such as \code{\{"id": 17, "takeaway": "unrelated content"\}} to replace the segment with a short takeaway. However, based on our preliminary experiments, even with powerful LLMs such as Claude 4 Sonnet and Gemini 2.5 Pro, the agent often persisted with the original task without calling \code{erase}, despite explicit prompts. Below is an example from the \code{django__django-13012} instance in SWE-bench Verified, where the Claude 4 Sonnet model insisted on continuing the original task even when prompted aggressively and explicitly for trajectory reduction:

\begin{infobox}{A failing example \texttt{(django\_\_django-13012)}}
    \textbf{[System]} \ldots{} The user will request for reflection on previous steps by saying "\#reflection". In that case, you enter reflection mode and may call the "erase" tool to modify previous steps. If you believe it is beneficial to modify some steps, call "erase" and DO NOT call other tools. Otherwise, DO NOT continue the task and DO NOT call any tool.

    \vspace{.5em}
    \textbf{[User]} Problem statement: \ldots{}

    \vspace{.5em}
    \textbf{[Assistant]} \ldots{}\enspace{} \textbf{[Tool]} \ldots{}\enspace{} \textbf{[Assistant]} \ldots{}\enspace{} \textbf{[Tool]} \ldots{}\enspace{} \textbf{[Assistant]} \ldots{}\enspace{} \textbf{[Tool]} \ldots{}
    
    \emph{(existing steps in the trajectory)}
    
    \vspace{.5em}
    \textbf{[User]} \#reflection. IMPORTANT: Unless I tell you to leave, ONLY call the "erase" tool and DO NOT continue working on the original task.

    \vspace{.5em}
    \sethlcolor{magenta!30}
    \textbf{[Assistant]} I'm examining the Django codebase to find how ExpressionWrapper and GROUP BY clauses interact. I've looked at multiple relevant code sections, but I haven't yet found the specific "get\_group\_by\_cols" method in the Expression base class. Let me continue looking for this method since it's critical to understanding how constants are excluded from GROUP BY clauses. \hl{Let me continue by examining more of the Expression class code to find the relevant "get\_group\_by\_cols" method: \ldots{} \emph{(calling other tools)}}
\end{infobox}

A possible reason is that the LLM has memorized the standard procedure of program repair during training, leading to an uncontrollable tendency to follow that procedure given the context. Therefore, enabling the agent to reduce its own trajectory would require careful fine-tuning, which is resource-intensive, error-prone, and often impossible for proprietary LLMs.

To overcome this problem, we took a step back and moved the trajectory reduction step to a separate module called the \emph{reflection module}. This LLM-based module is explicitly called by the outer system to reduce the trajectory when there is an estimated benefit. In this way, the timing of trajectory reduction is controlled by an external system, and the the LLM that runs the agent remains unaware of the reduction process, minimizing disturbance to its original workflow.

\subsubsection{Controlling the Overhead}
\label{sec:motivation-design-overhead}

In terms of efficiency, the current design with a separate reflection module is not ideal due to the additional cost overhead. Passing the full agent trajectory to the reflection module at each step would double the token usage, defeating the whole purpose of cost reduction. Making the situation worse, the KV Cache cannot be shared between the target agent and the reflection module due to differences in the system prompts and tools. Furthermore, modifying a step in the trajectory invalidates all subsequent caches. Therefore, controlling the overhead becomes an essential topic.

The first approach to consider is the choice of LLM in the reflection module. Since trajectory reduction is arguably easier than the agent's primary task, we can significantly reduce overhead by choosing a more cost-efficient LLM for the reflection module. In the case study illustrated in \figref{motivation-casestudy}, the trajectory was generated by the agent using Claude 4 Sonnet, while the reduction was performed by GPT-5 mini, which is 12 times cheaper than Claude 4 Sonnet, minimizing the overhead concern. We will discuss more about the choice of LLM in \secref{impl-hyperparam}.

\begin{figure}
    \centering
    \includegraphics[width=1\linewidth]{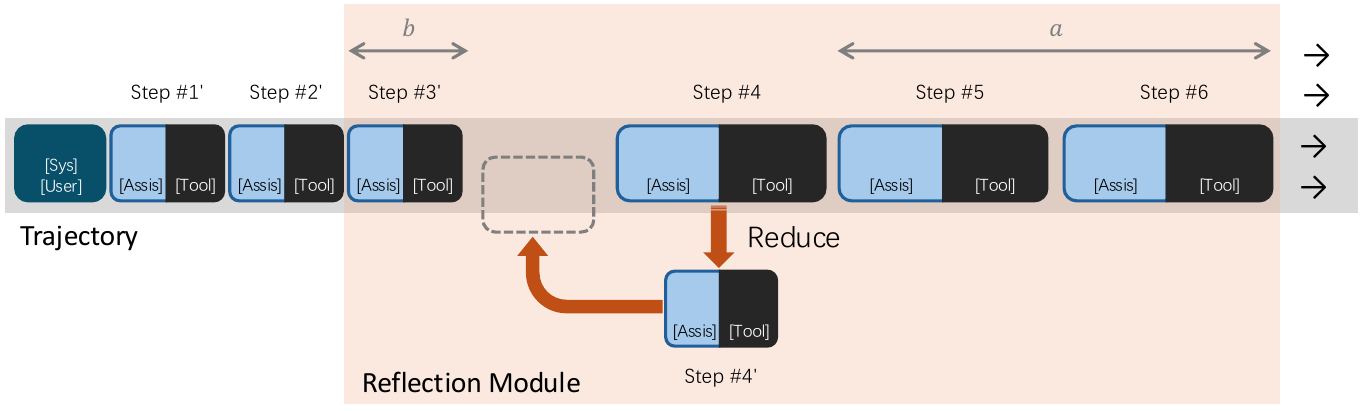}
    \caption{Design of the reflection module in \ourtool.}
    \label{fig:motivation-design}
\end{figure}

To further reduce overhead, we reduce the amount of context provided to the LLM used in the reflection module. We propose a sliding-window approach, as shown in \figref{motivation-design}. The orange box in the figure refers to the reflection module. When the agent reaches step $s$, the reflection module is allowed to reduce only the content in step $s-a$, and given only a fixed amount of context, from step $s-a-b$ to step $s$. The module prompts a cost-efficient LLM to reduce the content in step $s-a$ by removing useless, redundant, and expired information with the given context. $a$ and $b$ are hyperparameters that can be adjusted to balance the cost overhead and agent performance caused by trajectory reduction. Furthermore, we skip the trajectory reflection process if the length of step $s-a$ is within a token threshold $\theta$, where the benefit would be small. \figref{motivation-design} illustrates the approach where $a=2$, $b=1$ and $\theta=0$. The choice of these hyperparameters will be discussed in \secref{impl-hyperparam}.

From the perspective of efficiency, the token usage of each trajectory reduction step is capped at the length of $a+1+b$ steps. Additionally, since it only modifies a fixed recent step (step $s-a$), KV Cache entries corresponding to previous steps are preserved, minimizing overhead from cache invalidation. From the perspective of agent performance, this design ensures that the reflection module cannot destructively erase the most recent step or all steps at once, preventing disastrous outcomes from occasional LLM failures.

%% file: sections/3_approach.tex
\section{Algorithm}
\label{sec:approach}

In \secref{motivation-design} and \figref{motivation-design}, we have established the design of \ourtool with three technical insights:

\begin{itemize}
    \item Prompting a cost-efficient LLM with the three types of waste is enough to achieve an effective trajectory reduction.
    \item Manually invoking the reflection module after a long step is necessary, because agents are incapable of using the erase tool autonomously.
    \item The sliding window balances the overhead while retaining reduction effectiveness.
\end{itemize}

\newcommand*{\tikzmk}[1]{\tikz[remember picture,overlay,] \node (#1) {};\ignorespaces}
\newcommand{\boxit}[1]{\tikz[remember picture,overlay]{\node[fill=#1,opacity=.15,fit={($(A)+(0,-.5\baselineskip)$)($(B)+(.88\linewidth,0)$)}] {};}\ignorespaces}

\begin{algorithm}[htb]
    \caption{Integrating \ourtool in a typical LLM agent}
    \label{alg:approach}

    \begin{algorithmic}[1]
        \Statex \textbf{Input:} Problem instruction ($I$), Environment ($E$)
        \Statex \textbf{Output:} Result ($r$), Modified environment ($E$)
        \Statex \textbf{Constant:} Number of steps before ($b$) and after ($a$) the target step in context, Step limit ($s_\text{max}$), Length threshold ($\theta$), LLM for agent steps ($\text{LLM}_\text{agent}$) and reflection steps ($\text{LLM}_\text{reflect}$)

        \Statex

        \State $T \gets \textsc{MakeInitialPrompt}(I)$
            \Comment{Initialize the trajectory}
            \label{line:approach:trivial-begin}

        \ForEach{$s \in [1 \ldots s_\text{max}]$}
            \State $m_\text{assis} \gets \text{LLM}_\text{agent}(T)$
                \Comment{Perform agent step}
        
            \If{$\textsc{IsTaskDone}(m_\text{assis})$}
                \State $r \gets \texttt{"finished"}$
                \State \Return r, E
            \EndIf

            \State $E, m_\text{tool} \gets \textsc{ExecTool}(E, m_\text{assis})$
                \Comment{Parse and execute the tool call in the environment}
            \State $T \gets T + [\langle m_\text{assis}, m_\text{tool} \rangle]$
                \Comment{Concatenate messages in this step into the trajectory}
                \label{line:approach:trivial-end}

            \Statex

            \tikzmk{A}\If{$s-a > 0$} \label{line:approach:reflection-begin}
                \State $l_\text{orig} \gets \textsc{Length}(\textsc{Serialize}([T[s-a]]))$
                    \label{line:approach:calc-len}
                \If{$l_\text{orig} > \theta$}
                        \label{line:approach:len-threshold-in}
                    \State $ctx \gets \textsc{Serialize}(T[\text{max}(0, s-a-b) : s])$
                    \State $m_\text{reduced} \gets \text{LLM}_\text{reflect}(ctx, s-a)$
                        \Comment{Perform reflection step}
                    \State $l_\text{reduced} \gets \textsc{Length}(m_\text{reduced})$

                    \Statex
                    
                    \If{$l_\text{orig} - l_\text{reduced} > \theta$}
                            \label{line:approach:len-threshold-out}
                            \Comment{Apply reduction if benefit reaches the threshold}
                        \State $T[s-a] \gets m_\text{reduced}$
                            \label{line:approach:apply-reduction}
                    \EndIf
                \EndIf
            \EndIf \label{line:approach:reflection-end}
            \tikzmk{B}\boxit{orange!80}

        \Statex
        
        \EndFor

        \Statex

        \State $r \gets \texttt{"interrupted"}$
            \Comment{Reaching the step limit}
        \State \Return r, E
    \end{algorithmic}
\end{algorithm}

Here we instantiate this design into a concrete algorithm, as shown in \algoref{approach}.

As a typical LLM agent, \algoref{approach} takes two inputs: the problem instruction $I$, which contains a natural language description of the task, and the environment $E$, which the agent can interact with through tool calls. The output is the result $r$, which indicates whether the task is completed. The environment $E$ is modified by the agent in place. The algorithm also contains several constants that are adjustable hyperparameters.

The first part (lines \ref{line:approach:trivial-begin}-\ref{line:approach:trivial-end}) of this algorithm is directly based on the workflow of existing LLM agents. It first initializes the trajectory $T$ with only the initial system and user prompts constructed with $I$. Then, it iteratively calls the $\text{LLM}_\text{agent}$ function to generate an assistant message ($m_\text{assis}$) that contains the next action to take based on $T$. If the action is to finish the task, the agent stops; otherwise, the \textsc{ExecTool} function interacts with the environment $E$ based on the tool call and retrieves the result as $m_\text{tool}$. The messages generated in this step ($m_\text{assis}$ and $m_\text{tool}$) are then concatenated into $T$.

Lines \ref{line:approach:reflection-begin}-\ref{line:approach:reflection-end} highlighted by the orange box represent the reflection module introduced in \ourtool. As illustrated in \figref{motivation-design}, the reflection module employs a sliding window and aims to reduce the length of step $s-a$ when the agent reaches step $s$. It uses another language model, $\text{LLM}_\text{reflect}$, which can be more cost-efficient than $\text{LLM}_\text{agent}$ to reduce the overhead of trajectory reduction. The module is controlled by the hyperparameter $\theta$, which is the minimum number of tokens required for a beneficial reduction, and by $a$ and $b$, which define the number of steps included in the context given to $\text{LLM}_\text{reflect}$. It first serializes the target step into a string (\textsc{Serialize}) and calculates its tokenized length (\textsc{Length}) as $l_\text{orig}$ at line \ref{line:approach:calc-len}. If the length is below the threshold ($l_\text{orig} \leq \theta$), reflection is skipped because the benefit may not exceed the overhead. Otherwise, it constructs the context ($ctx$) and asks $\text{LLM}_\text{reflect}$ to generate a reduced version of the target step based on $ctx$. If the length reduction exceeds the token threshold $\theta$, the target step in the trajectory $T$ is replaced by the reduced version at line \ref{line:approach:apply-reduction}.

\section{Implementation}
\label{sec:impl}

\subsection{Integration in Agents}
\label{sec:impl-agent}

\ourtool is a general approach that can be readily applied to any LLM agent fitting the definition ``agents are models using tools in a loop''. For a typical agent system similar to the example in \algoref{approach}, we can add a call to the reflection module (highlighted by the orange box) after each step in the agentic loop.

Following this procedure, we integrated \ourtool into Trae Agent, which is open-source and ranked first on SWE-bench Verified at the time of our research (July 2025). We did not further integrate or experiment with \ourtool in other LLM agents, based on the observation that current agent systems are generally homogeneous and employ a similar set of prompts and tools. For example, Trae Agent is equipped with four tools: \code{bash} to execute Bash commands, \code{str_replace_editor} to view or edit files, \code{think} to analyze the issue, and \code{task_done} to finish the task. This set of tools is semantically similar to those in other systems, such as mini-SWE-agent~\cite{yang2024sweagent} and OpenHands~\cite{wang2024openhands}. Therefore, the results with Trae Agent are likely to generalize to a variety of similar agent systems.

Note that there are variants of agent systems, such as \emph{ensembled} systems that call the LLM agent multiple times and decide the final answer by majority voting or an LLM, and \emph{multi-agent} systems that simultaneously spawn multiple LLM agents with a communication mechanism. \ourtool can be integrated into such systems by adding the reflection module to the LLM agent in an ensembled system, or to all LLM agents in a multi-agent system. To maintain the generality of the approach, this paper does not focus on optimizations specific to these systems. We leave the evaluation of such applications for future work.

\subsection{Hyperparameter Settings}
\label{sec:impl-hyperparam}

In this subsection, we discuss the effect of different hyperparameter settings ($\text{LLM}_\text{reflect}$, $\theta$, $a$, and $b$, as shown in \algoref{approach}) based on the Trae Agent integration. Note that the other two hyperparameters ($\text{LLM}_\text{agent}$ and $s_\text{max}$) already exist in the original Trae Agent, so we exclude them from the discussion and set them to standard default values: $\text{LLM}_\text{agent} =$ Claude 4 Sonnet, and $s_\text{max} =$ 50.

\subsubsection{The Process}

Since the search space of hyperparameters is multidimensional, we used the standard hill climbing algorithm to find an ideal setting, which is arguably more rigorous than determining each parameter independently.

The process began by setting each hyperparameter to an initial default value. Then, we quantitatively compared efficiency and agent performance via a small experiment involving variants. Each variant adjusted one hyperparameter to a different value in the search space while keeping other hyperparameters at their default values. If a variant demonstrated a better overall result, it was selected as the new default value for that hyperparameter in the next iteration. The process repeated until the hyperparameters converged to a local optimum.

\subsubsection{Benchmark}

In this process, we used a random subset of 100 instances from SWE-bench Verified~\cite{jimenez2024swebench} as the benchmark for deciding hyperparameters. This was the same set used in the manual analysis in \secref{motivation-waste-examples}. We did not use the full benchmark because the remaining instances will be used exclusively for evaluation in \secref{evaluation}.

\subsubsection{The Search Space}
\label{sec:impl-hyperparam-llm}

\begin{table}[tb]
    \centering
    \caption{Compared LLMs and baselines with their pricing information.}
    \label{tab:approach-llms-pricing}

\begin{tabular}{rl|rrr}
\toprule
\multicolumn{1}{l}{}       &                    & \multicolumn{3}{c}{Pricing (US\$ / M tokens)} \\
\multicolumn{1}{l}{}       &                    & Cached Input       & Input       & Output     \\
\midrule
\multirow{5}{*}{LLMs}      & Claude 3.5 Haiku   & 0.080              & 0.800       & 4.000      \\
                           & Gemini 2.5 Flash   & 0.075              & 0.300       & 2.500      \\
                           & GPT-5 mini         & 0.030              & 0.250       & 2.000      \\
                           & DeepSeek v3 (0324) & 0.070              & 0.270       & 1.100      \\
                           & Qwen 3 (2507)      & 0.070              & 0.270       & 1.100      \\
\midrule
\multirow{2}{*}{Baselines} & LLMLingua-2        & 0.010 & 0.010 & Free \\
                           & Original / Random / Delete & Free & Free & Free \\
\bottomrule
\end{tabular}
\end{table}

We considered five variants for $\text{LLM}_\text{reflect}$, the LLM used in the reflection module. These variants are selected to cover a variety of cost-efficient LLMs from various vendors: Claude 3.5 Haiku from Anthropic, Gemini 2.5 Flash from Google, GPT-5 mini from OpenAI, DeepSeek v3 (0324) from DeepSeek, and Qwen 3 (2507) from Alibaba. The pricing information for these LLMs is listed in \tabref{approach-llms-pricing}, which was used to calculate their computational cost.

As a sanity check, we also compared with the following four additional baseline variants:

\begin{itemize}
    \item \textbf{LLMLingua-2}~\cite{pan2024llmlingua2}: A small model for prompt compression. Since this model is hosted locally, we estimated its pricing (as US\$0.01 per million input tokens) based on a similar BERT model with 600M parameters.
    \item \textbf{Original}: The unmodified agent system that skips the reflection step and deletes nothing.
    \item \textbf{Random}: A baseline that randomly deletes 75\% of processed tokens in the reflection step.
    \item \textbf{Delete}: A baseline that deletes all processed tokens in the reflection step.
\end{itemize}

Note that we used the term \emph{processed} tokens for the Random and Delete baselines, because $\text{LLM}_\text{reflect}$ only processes a step when its length is greater than $\theta$ tokens after a delay of $a$ steps. Therefore, even the Delete baseline will retain all tokens in a short or recent step.

For the other three hyperparameters ($\theta$, $a$, and $b$), we considered these variants:

\begin{itemize}
    \item $\theta =$ 0, 250, 500, 1000, 2000.
    \item $a =$ 0, 1, 2, 3.
    \item $b =$ 0, 1, 2.
\end{itemize}

Note that we limited the choice of $a$ and $b$ to an upper bound ($a \leq 3, b \leq 2$) because higher values would strictly increase cost overhead and potentially improve agent performance. Since the variants of $a=2$ and $b=1$ already exhibit satisfactory performance, as will be shown in the results, we estimated that there would be minimal benefit in further increasing $a$ or $b$.

\subsubsection{Metrics}
\label{sec:impl-hyperparam-metrics}

We used the following metrics to measure the amount of \textbf{efficiency improvement} for each variant:

\begin{itemize}
    \item \textbf{Keep\%}: The average percentage of trajectory tokens retained by the LLM in the reflection module, formally $\sum l_\text{reduced} \:/\: \sum l_\text{orig} * 100$ in \algoref{approach}. For example, the Delete baseline has Keep\% = 0, and the Original baseline has Keep\% = 100. 
    \item \textbf{I} and \textbf{O}: Accumulated usage of input (``prompt'') or output (``completion'') tokens. For easier comparison, we normalized all numbers in these two rows so that the Original baseline has $I=1$.
    \item \textbf{\$} and \textbf{\$+}: LLM cost for agent steps (\$) and reflection steps (\$+). This value is calculated based on the pricing for the LLMs involved. The cost covers both input tokens (considering the discount for tokens in KV Cache) and output tokens. We normalized all numbers in these two rows so that the Original baseline has $\$=1$. We separately reported the cost for agent steps and reflection steps to  help readers better understand the overhead caused by the reflection module.
\end{itemize}

Additionally, we used the following metrics to measure the impact on the \textbf{agent's performance}:

\begin{itemize}
    \item \textbf{Pass\%}: The percentage of successfully resolved instances in the benchmark.
    \item \textbf{Step} and \textbf{PStep}: The average number of total agent steps for all instances (Step) and for only successfully resolved instances (PStep). Lower is better because an increase in steps indicates that some information is inappropriately reduced, disturbing the agent and requiring it to recover the information with additional steps.
\end{itemize}

\subsubsection{Results}

We set the initial hyperparameters as $\text{LLM}_\text{reflect} =$ Gemini 2.5 Flash, $\theta=500$, $a=3$, and $b=1$. During the first iteration, two variants (one with $\text{LLM}_\text{reflect} =$ GPT-5 mini, the other with $a=2$) showed better overall results, so we updated the values of these two hyperparameters. During the second iteration, no variant was better than the current setting, indicating that the hyperparameters had converged. Below, we report the results of the second iteration.

\begin{table}[tb]
    \centering
    \caption{Results of different LLMs or baselines as $\text{LLM}_\text{reflect}$.}
    \label{tab:impl-hyperparam-result-llms}

\begin{tabular}{c|rrrrr|rrrr}
\toprule
       & Claude & Gemini & GPT    & DeepSeek & Qwen   & Orig. & Lingua & Random & Delete \\
\midrule
Keep\% & 14.4   & 21.9   & 28.6   & 23.7     & 29.0   &          & 22.5   & 25.0   & 0 \\
I      & 0.473  & 0.553  & 0.586  & 0.606    & 0.722  & 1.000    & 0.603  & 0.642  & 0.428 \\
O      & 0.012  & 0.012  & 0.012  & 0.012    & 0.012  & 0.012    & 0.013  & 0.013  & 0.013 \\
\$     & 0.652  & 0.698  & 0.707  & 0.733    & 0.818  & 1.000    & 0.753  & 0.794  & 0.655 \\
\$+    & 0.148  & 0.078  & 0.077  & 0.052    & 0.065  &          & 0.000  &        & \\
\midrule
Pass\% & 60     & 63     & 65     & 64       & 62     & 65       & 61     & 64     & 58 \\
Step   & 42.39  & 40.95  & 38.90  & 40.41    & 41.10  & 39.74    & 43.89  & 44.95  & 45.43 \\
PStep  & 40.27  & 39.46  & 37.34  & 38.47    & 39.69  & 38.29    & 42.21  & 43.66  & 43.69 \\
\bottomrule
\end{tabular}
\end{table}

\tabref{impl-hyperparam-result-llms} shows the results for the $\text{LLM}_\text{reflect}$ hyperparameter. The first row (Keep\%) shows that all five LLMs can remove most of the content in the trajectory steps where the length is at least $\theta = 500$ tokens. However, different LLMs have different characteristics, keeping 14.4\%--29.0\% of tokens in the processed steps. As a result, the accumulated input tokens (I) drop to 47.3\%--72.2\%, while the output tokens (O) remain close to the Original baseline. The agent cost (\$) drops to 65.2\%--81.8\% of Original. The ``\$+'' row indicates that $\text{LLM}_\text{reflect}$ introduces an overhead cost of 5.2\%--14.8\%, depending on the model used.

The last three rows measure the performance impact of trajectory reduction on the SWE-bench Verified instances. While the numbers in Pass\% vary between LLMs and baselines, all variants reach a Pass\% of at least 58\% (for the Delete baseline), which is close to the 65\% achieved by the Original agent. It may be surprising that even trivial baselines such as Delete and Random have reasonably well agent performance. The reason is that the choice of other hyperparameters ($a=2$ and $\theta=500$ in this experiment) ensures that $\text{LLM}_\text{reflect}$ can only manipulate steps of excessive length after a delay of 2 steps. Therefore, the disruption to $\text{LLM}_\text{agent}$ is limited, and the LLM can recover the disrupted information autonomously through additional tool calls. This leads to an increase in ``Step'' and ``PStep'' of more than 13\%, but not a large drop in ``Pass\%''. In other words, \ourtool protects against a malfunctioning $\text{LLM}_\text{reflect}$ and prevents catastrophic failures. We can see that the increase is most severe for the Delete baseline, which maximizes the disruption to the trajectory. Such baselines are generally impractical in production due to longer task completion times and the increased risk of instability.

Based on the results, GPT-5 mini maintains the same Pass\% as Original, and is also the only model with a decrease in Step and PStep, demonstrating that it does not compromise the agent's performance while saving significant token costs. Therefore, GPT-5 mini was chosen as the final $\text{LLM}_\text{reflect}$ in \ourtool.

\begin{table}[tb]
    \centering
    \caption{Results of different hyperparameter settings ($\theta$, $a$, and $b$).}
    \label{tab:impl-hyperparam-result-threshold}

\begin{tabular}{c|r|rrrrr}
\toprule
       &       & \multicolumn{5}{c}{$\theta$}               \\
       & Orig. & 0      & 250    & 500    & 1000   & 2000   \\
\midrule
Keep\% &       & 32.6   & 32.0   & 28.6   & 24.3   & 16.2   \\
I      & 1.000 & 0.547  & 0.587  & 0.586  & 0.662  & 0.728  \\
O      & 0.012 & 0.011  & 0.011  & 0.012  & 0.012  & 0.012  \\
\$     & 1.000 & 0.669  & 0.700  & 0.707  & 0.765  & 0.816  \\
\$+    &       & 0.118  & 0.100  & 0.077  & 0.040  & 0.017  \\
\midrule
Pass\% & 65    & 62     & 65     & 65     & 60     & 58     \\
Step   & 39.74 & 38.57  & 39.68  & 38.90  & 39.70  & 40.22  \\
PStep  & 38.29 & 37.40  & 38.09  & 37.34  & 38.27  & 38.91  \\
\bottomrule
\end{tabular}

\vspace{.5em}

\begin{tabular}{c|rrrr|rrr}
\toprule
       & \multicolumn{4}{c|}{$a$}          & \multicolumn{3}{c}{$b$}  \\
       & 0      & 1      & 2      & 3      & 0      & 1      & 2      \\
\midrule
Keep\% & 31.7   & 31.0   & 28.6   & 31.3   & 31.5   & 28.6   & 34.5   \\
I      & 0.727  & 0.569  & 0.586  & 0.624  & 0.644  & 0.586  & 0.719  \\
O      & 0.013  & 0.011  & 0.012  & 0.012  & 0.012  & 0.012  & 0.011  \\
\$     & 0.843  & 0.688  & 0.707  & 0.736  & 0.749  & 0.707  & 0.790  \\
\$+    & 0.081  & 0.067  & 0.077  & 0.078  & 0.075  & 0.077  & 0.078  \\
\midrule
Pass\% & 59     & 62     & 65     & 66     & 64     & 65     & 65     \\
Step   & 44.94  & 39.13  & 38.90  & 40.25  & 40.31  & 38.90  & 39.32  \\
PStep  & 43.59  & 37.73  & 37.34  & 40.06  & 39.45  & 37.34  & 37.49  \\
\bottomrule
\end{tabular}

\end{table}

The results for the remaining hyperparameters, $\theta$, $a$, and $b$, are illustrated in \tabref{impl-hyperparam-result-threshold}. For $\theta$, a higher threshold value results in more input tokens for the agent (I), but reduced reflection overhead. $\theta=500$ balances token savings and reflection overhead. In terms of agent performance, $\theta=500$ also demonstrates optimal performance close to that of Original.

For $a$ and $b$, a higher value delays the trajectory reduction and gives the reflection module more context, hence resulting in better performance but reduced efficiency. $a=2, b=1$ is the minimum choice of $a$ and $b$ that has a negligible impact to performance, while still improving the efficiency by more than 22\% as shown in the \$ and \$+ rows.

%% file: sections/4_eval.tex
\section{Evaluation}
\label{sec:evaluation}

In this section, we empirically evaluate \ourtool with the following research questions.

\begin{enumerate}[left=1em]
    \item[RQ1.] \textbf{Efficiency Improvement:} Can \ourtool improve the efficiency of LLM agents by reducing the length of the trajectory?
    \item[RQ2.] \textbf{Performance Impact:} Does \ourtool harm the task-solving performance of LLM agents?
    \item[RQ3.] \textbf{Generalization:} How do the results generalize to different benchmarks and LLMs?
\end{enumerate}

\subsection{Experimental Setup}

\subsubsection{Experiment Subjects}

We use the \ourtool integrated into Trae Agent~\cite{gao2025trae} as discussed in \secref{impl}. To assess the generalization of our approach (RQ3), we evaluate the Trae Agent integration with two recent LLMs (i.e., $\text{LLM}_\text{agent}$) that are common for coding agents: Claude 4 Sonnet and Gemini 2.5 Pro. For each LLM, we compare the results of \ourtool against the Original baseline, where the reflection step is skipped.

\subsubsection{Metrics}

We use the same set of metrics as previously described in \secref{impl-hyperparam-metrics}, which contains five metrics for efficiency (Keep\%, I, O, \$, and \$+), and three metrics for performance impact (Pass\%, Step, and PStep). For RQ1, we can infer that \ourtool improves efficiency if Keep\% is low and I/O/\$ decrease over Original. For RQ2, we can infer that \ourtool does not harm the agent's performance if Pass\% does not decrease and Step/PStep does not increase over Original.

\subsubsection{Benchmarks}

We use two benchmarks for the evaluation. \textbf{(1) SWE-bench Verified}~\cite{jimenez2024swebench} contains 500 human-verified software engineering tasks in Python based on real-world GitHub issues. Since 100 instances in this benchmark are used for manual analysis (in \secref{motivation-waste-examples}) and hyperparameter selection (in \secref{impl-hyperparam}), we exclude them from the experiment in this section and randomly select 200 instances from the remaining 400 instances. Similar to the validation/test split in standard machine learning practice, this experimental setup mitigates the risk of hyperparameter overfitting. The list of selected instances is available in the artifact.
\textbf{(2) Multi-SWE-bench Flash}~\cite{zan2025multi} contains 300 instances based on GitHub issues, similar to SWE-bench Verified. However, this benchmark covers tasks in seven other programming languages, comprising 45 Rust instances, 45 TypeScript instances, 45 JavaScript instances, 40 Java instances, 45 Go instances, 40 C instances, and 40 C++ instances. These tasks are generally harder than SWE-bench Verified, and often require the agent to figure out how to build the project, whereas SWE-bench Verified provides the agent with a fully working environment from the start.

\subsubsection{Hyperparameters}

For the hyperparameters related to \ourtool, we follow the experiment results in \secref{impl-hyperparam}, setting $a=2$, $b=1$, $\theta=500$ and $\text{LLM}_\text{reflect} = \text{GPT-5 mini}$. For the step limit ($s_\text{max}$), we set it to 50 for the SWE-bench Verified benchmark as the default value in Trae Agent. $s_\text{max}$ is increased to 100 for Multi-SWE-bench Flash because this benchmark requires more steps to complete, as will be shown from the Step/PStep metrics in the experiment results.

\subsection{Results}

\tabref{evaluation-result} contains the main results of the experiment, and we discuss them with respect to the three research questions, i.e., efficiency improvement, performance impact, and generalization.

\begin{table}[tb]
    \centering
    \caption{Efficiency improvement and agent performance impact of \ourtool.}
    \label{tab:evaluation-result}

\begin{tabular}{c|rr|rr|rr|rr}
\toprule
Benchmark & \multicolumn{4}{c|}{SWE-bench Verified}                                  & \multicolumn{4}{c}{Multi-SWE-bench Flash}                               \\
\midrule
$\text{LLM}_\text{agent}$ & \multicolumn{2}{c|}{Claude 4 Sonnet} & \multicolumn{2}{c|}{Gemini 2.5 Pro} & \multicolumn{2}{c|}{Claude 4 Sonnet} & \multicolumn{2}{c}{Gemini 2.5 Pro} \\
Approach  & \small Orig. & \small \ourtool & \small Orig. & \small \ourtool & \small Orig. & \small \ourtool & \small Orig. & \small \ourtool \\
\midrule
Keep\%    &                 & 30.8              &                 & 22.6             &                 & 30.2              &                & 25.7              \\
I         & 1.000           & 0.601             & 1.000           & 0.591            & 1.000           & 0.596             & 1.000          & 0.403             \\
O         & 0.012           & 0.011             & 0.007           & 0.006            & 0.006           & 0.006             & 0.007          & 0.009             \\
\$        & 1.000           & 0.714             & 1.000           & 0.623            & 1.000           & 0.676             & 1.000          & 0.559             \\
\$+       &                 & 0.074             &                 & 0.118            &                 & 0.055             &                & 0.082            \\
\midrule
Pass\%    & 64.5            & 66.5              & 50.5            & 52.0             & 40.0            & 39.0              & 21.7           & 22.7              \\
Step      & 39.62           & 39.95             & 37.98           & 37.44            & 70.37           & 70.45             & 57.20          & 43.90             \\
PStep     & 37.75           & 38.21             & 32.59           & 31.72            & 62.08           & 60.77             & 37.86          & 29.75             \\
\bottomrule
\end{tabular}
\end{table}

\subsubsection{RQ1. Efficiency Improvement}

\begin{figure}
    \centering
    \includegraphics[width=1\linewidth]{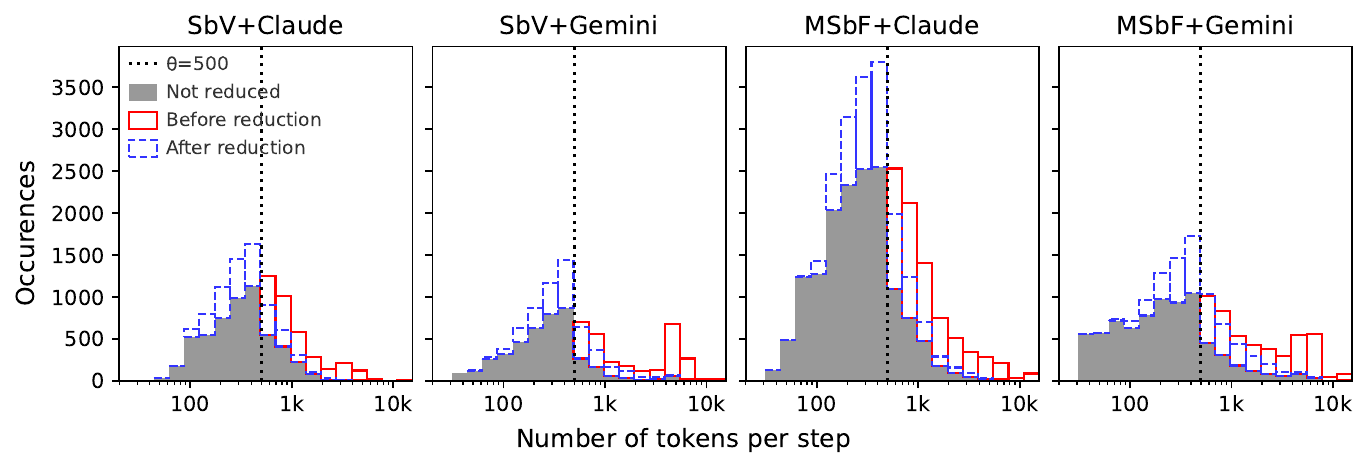}
    \caption{Histogram of the number of tokens reduced by \ourtool in each step. The X axis is plotted logarithmically. SbV = SWE-bench Verified. MSbF = Multi-SWE-bench Flash.}
    \label{fig:evaluation-length}
\end{figure}

We analyze how the trajectory reduction by \ourtool improves efficiency through various metrics reported in \tabref{evaluation-result}. First, the reflection module removes 69.2\%--77.4\% of the content it processes (calculated by 1 - $\frac{\text{Keep\%}}{100}$). This results in a reduction in the accumulated input tokens (1 - I) of 39.9\%--59.7\%. This reduction is lower than the percentage of deleted tokens among the processed tokens because the reflection module only processes steps longer than the token threshold ($\theta = 500$) and only applies the reduction after $a=2$ steps. The reduced input tokens then lead to a reduction in the computational cost of the agent (\$) of 28.6\%--44.1\%. This reduction is smaller due to costs related to output tokens and invalidated KV Caches, but is still significant. After we further consider the computational overhead of the reflection module itself (\$+), the final cost reduction becomes 21.1\%--35.9\%.

\figref{evaluation-length} further visualizes the reduction of tokens in each step. In this histogram, the gray area represents the steps not reduced by \ourtool due to two possible reasons: (1) For the area on the left of the vertical line representing $\theta = 500$ tokens, reflection is skipped to avoid the overhead of reducing short steps, as described by line \ref{line:approach:len-threshold-in} in \algoref{approach}; (2) For the area on the right of the line, $\text{LLM}_\text{reflect}$ is invoked but the reduction is not applied to the trajectory, as described by line \ref{line:approach:len-threshold-out} in \algoref{approach}. This mechanism minimizes the overhead of invoking $\text{LLM}_\text{reflect}$ and invalidating the KV Cache of the target agent. The red and blue squares, drawn on top of the gray area, represent the steps actually reduced by \ourtool. We observe that the reduced steps (blue) have significantly fewer tokens than their original content (red) in this logarithmic plot, indicating that the prompt in \ourtool can effectively detect and remove a significant amount of waste in such steps.

Note that the numbers in ``\$'' and ``\$+'' rows are normalized to facilitate relative comparison with Original. If we convert the numbers back to the average US\$ cost per instance, the Original baseline costs \$0.535, \$0.385, \$1.277, and \$0.701 for SbV+Claude, SbV+Gemini, MSbF+Claude, and MSbF+Gemini, respectively. \ourtool decreases the cost to \$0.422, \$0.285, \$0.933, and \$0.449, respectively. Considering that each instance only represents a single minor task of a software engineer (with a typical code modification in one or a few functions), the total savings among active users of a popular AI product can be substantial.

\begin{rqbox}
    \textbf{Finding 1.} \ourtool significantly removes waste in the trajectory, leading to a reduction in input tokens by 39.9\%--59.7\%, or a reduction in the final computational cost by 21.1\%--35.9\%, compared to the Original baseline.
\end{rqbox}

\subsubsection{RQ2. Performance Impact}

From the ``Pass\%'' row in \tabref{evaluation-result}, we can compare the numbers between Original and \ourtool to assess its impact on the agent's performance. On two benchmarks and two LLMs, the performance of \ourtool is comparable (-1.0\%--+2.0\%) to that of Original. These results show that \ourtool does not harm the agent's performance while improving efficiency, which contradicts the common belief of ``test-time compute''~\cite{beeching2024scaling} that there is a trade-off between token efficiency and model performance. As discussed in \secref{motivating-preliminary}, a possible explanation is that the performance of LLMs degrades as the length of the context increases~\cite{li2025longcodeu} or when the context is of low quality~\cite{laban2025llms}. Therefore, the removal of waste information from the agent trajectory may cause less degradation to the model's performance.

\begin{figure}
    \centering
    \includegraphics[width=1\linewidth]{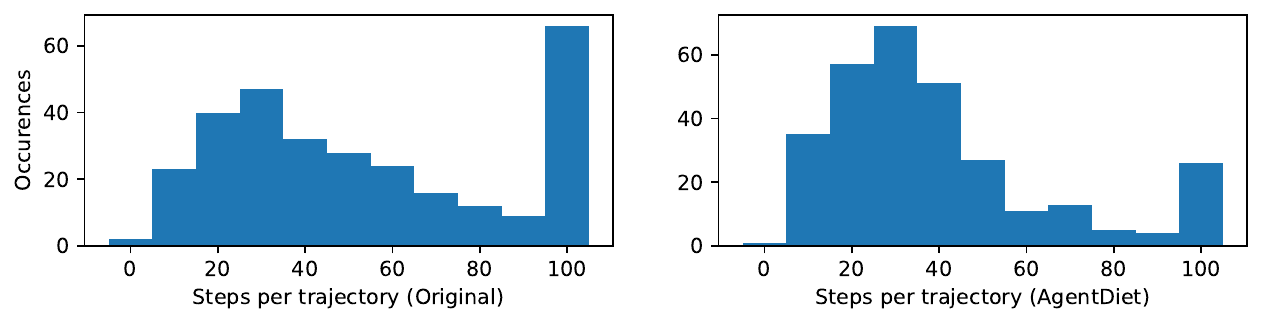}
    \caption{Histogram of the Step metric on Multi-SWE-bench Flash + Gemini 2.5 Pro between Original (left) and \ourtool (right).}
    \label{fig:evaluation-steps}
\end{figure}

Additionally, from the ``Step'' and ``PStep'' rows, we observe that \ourtool does not cause the steps required to resolve a task to increase. This is an additional indicator confirming that \ourtool does not disturb the agent.

We see an unexpected case in the last column in \tabref{evaluation-result}, where \ourtool significantly reduces the average steps from 57.20 to 43.90 when Gemini 2.5 Pro works on the Multi-SWE-bench Flash benchmark. We identify the reason by inspecting the trajectories: Gemini 2.5 Pro shows increasingly abnormal behavior when the context is too long, often resulting in repeated invalid tool calls until the step limit is reached. Therefore, \ourtool reduces this probability by reducing the length of the trajectory by half. \figref{evaluation-steps} presents the histogram of the step distribution in the experiment with Gemini 2.5 Pro + Multi-SWE-bench Flash, where \ourtool reduces the number of instances that reach the step limit (100 steps) from 66 to 26. As a result, the average steps are significantly reduced. This interesting observation highlights a side effect of \ourtool: improving the robustness of LLM agents in corner cases.

\begin{rqbox}
    \textbf{Finding 2.} The ratio of successfully resolved instances for \ourtool is comparable (-1.0\%--+2.0\%) to that of the Original baseline. \ourtool also does not cause the agent to take more steps to resolve a task. Therefore, trajectory reduction does not harm the agent's performance.
\end{rqbox}

\subsubsection{RQ3. Generalization}

The generalization of \ourtool across benchmarks and LLMs can be confirmed by comparing the columns in \tabref{evaluation-result}, where \ourtool consistently improves efficiency and maintains the same performance in two benchmarks (SWE-bench Verified and Multi-SWE-bench Flash) and two LLMs (Claude 4 Sonnet and Gemini 2.5 Pro). The results show that the current implementation of \ourtool, based on prompts optimized for only a few test cases (\secref{motivation-design}) and hyperparameters selected from a small dataset of 100 instances (\secref{impl-hyperparam}), can generalize to a variety of tasks.

\begin{table}[tb]
    \centering
    \caption{Efficiency improvement and performance impact of \ourtool on Multi-SWE-bench Flash instances categorized by programming languages.}
    \label{tab:evaluation-bylang}

\newcommand{\STAB}[1]{\begin{tabular}{@{}c@{}}#1\end{tabular}}

\begin{tabular}{c|c|rrrrrrr}
\toprule
 & Lang   & Rust   & TypeScript & JavaScript & Java   & Go    & C      & C++    \\
 & \#Inst.& 45     & 45         & 45         & 40     & 45    & 40     & 40     \\
\midrule
\multirow{8}{*}{\STAB{\rotatebox[origin=c]{90}{Claude 4 Sonnet}}}
 & Keep\% & 28.1   & 29.9       & 29.8       & 31.7   & 34.0  & 24.7   & 32.7   \\
 & I      & 0.603  & 0.560      & 0.649      & 0.637  & 0.615 & 0.543  & 0.602  \\
 & O      & 0.006  & 0.006      & 0.010      & 0.007  & 0.006 & 0.005  & 0.006  \\
 & \$     & 0.680  & 0.635      & 0.806      & 0.723  & 0.646 & 0.620  & 0.679  \\
 & \$+    & 0.061  & 0.049      & 0.085      & 0.057  & 0.049 & 0.043  & 0.049  \\
 & $\Delta$Pass  & -2     &+1          &+1          &+1      & -3    & -1     & 0      \\
 & $\Delta$Step  & -0.27  & -1.11      &+3.11       & -1.32  &+0.13  &+0.40   & -0.53  \\
 & $\Delta$PStep & -3.67  &+0.67       &+1.24       &+2.30   & -4.84 & -2.00  & -1.38  \\
\midrule
\multirow{8}{*}{\STAB{\rotatebox[origin=c]{90}{Gemini 2.5 Pro}}}
 & Keep\% & 22.3   & 25.5       & 34.1       & 26.1   & 28.1  & 18.3   & 24.4   \\
 & I      & 0.310  & 0.525      & 0.459      & 0.330  & 0.656 & 0.443  & 0.298  \\
 & O      & 0.007  & 0.015      & 0.010      & 0.008  & 0.013 & 0.009  & 0.006  \\
 & \$     & 0.425  & 0.746      & 0.597      & 0.438  & 0.885 & 0.638  & 0.424  \\
 & \$+    & 0.072  & 0.105      & 0.083      & 0.081  & 0.114 & 0.097  & 0.055  \\
 & $\Delta$Pass  & 0      &+3          &+4          & -1     &+1     & -3     & -1     \\
 & $\Delta$Step  & -13.91 & -11.36     & -11.44     & -15.88 & -3.02 & -10.15 & -27.52 \\
 & $\Delta$PStep & -1.64  &+0.58       &+1.02       & -5.10  & -0.76 & -1.67  & -3.38  \\
\bottomrule
\end{tabular}
\end{table}

Since Multi-SWE-bench Flash contains instances in seven programming languages, we can further break down the numbers by programming language to demonstrate the generalization across programming languages. The results are shown in \tabref{evaluation-bylang}, where ``$\Delta$Pass'', ``$\Delta$Step'', and ``$\Delta$PStep'' show the difference in resolved instances, average steps for all instances, and average steps for resolved instances between \ourtool and Original (a positive value means that the number for \ourtool is higher). The findings reported in previous RQs remain valid for each programming language in \tabref{evaluation-bylang}, confirming the generalization of the results.

\begin{rqbox}
    \textbf{Finding 3.} The implementation of \ourtool based on a small dataset can generalize to two different benchmarks (SWE-bench Verified and Multi-SWE-bench Flash), two different LLMs (Claude 4 Sonnet and Gemini 2.5 Pro), and seven different programming languages.
\end{rqbox}

%% file: sections/5_discussion.tex
\section{Discussion}
\label{sec:discussion}

\subsection{Future Work}

\subsubsection{Improving the Latency of Agents}

In \algoref{approach}, \ourtool adds an additional reflection step to the agent loop, which introduces an increase in latency. This increase is expected to be minimal, as we use a cost-efficient model for $\text{LLM}_\text{reflect}$, and the latency is reduced for $\text{LLM}_\text{agent}$ due to the reduction in trajectory length. We do not quantitatively compare latency between \ourtool and the Original agent because the experiment relies on commercial LLM APIs, where latency is highly unstable due to server load. In latency-critical scenarios, \algoref{approach} can be modified to perform the reflection and agent steps in parallel. This modification would eliminate the latency increase, at the cost of reducing the context available to the reflection module (hyperparameter $a$) by one step.

\subsubsection{Exploring Different Designs for Trajectory Reduction}

This paper presents a preliminary study on the efficiency of LLM agents and proposes \ourtool as a simple yet effective approach for trajectory reduction. Moreover, we only consider off-the-shelf LLMs as $\text{LLM}_\text{reflect}$, incurring a cost overhead of 5\%--10\%. Future work could replace off-the-shelf LLMs with more efficient customized models to reduce this overhead.

For agent systems that route tasks of varying difficulty to different LLMs, it is also possible to automatically adjust the hyperparameters of \ourtool to balance efficiency and performance across tasks.

\subsection{Threats to Validity}

\subsubsection{Generalization Across Agents}

The main threat to external validity comes from the generalization of our findings. We employ two benchmarks and two LLMs to mitigate this threat. However, due to effort and cost constraints, \ourtool is implemented only for the Trae Agent. In fact, the existing experiments have already incurred an LLM cost of approximately US\$ 2000, making experimentation with additional agent systems financially prohibitive. As discussed in \secref{impl-agent}, current agent systems are generally homogeneous, employing a similar set of prompts and tools. Therefore, we believe this threat has a limited impact on our findings.

\subsubsection{Data Leakage in LLMs}

Another potential threat to internal validity is data leakage in LLMs. Since the experiment utilizes proprietary LLMs, benchmark instances may have been included in their training data. We have no control over their training process. There are existing studies on this threat that readers can refer to. For example, the authors of SWE-bench empirically compared the performance of multiple LLMs before and after their knowledge cutoff date and concluded that ``for most models there's little difference in performance before or after this date''~\cite{jimenez2024swebench}. Thus, we believe this threat does not significantly affect the validity of our findings.

To further mitigate this threat, we included Multi-SWE-bench Flash~\cite{zan2025multi}, a recently published benchmark collected by a third party and not mentioned in the technical report of any evaluated LLM.

\subsubsection{Correctness of Patches}

Both SWE-bench Verified and Multi-SWE-bench Flash validate the correctness of agent-generated patches using developer-written test cases. An instance is deemed successfully resolved if it passes all test cases. However, a patch that passes all test cases (``plausible'') may not be semantically equivalent to the ground truth patch written by the developer (``correct''). Consequently, overfitting to test cases poses a potential threat to validity for all test-based approaches, including ours. However, prior work~\cite{petke2024patch} suggests that overfitting is not a severe threat in program repair. Additionally, since each instance in SWE-bench Verified and Multi-SWE-bench Flash includes at least one held-out test case, the threat of overfitting is mitigated. Furthermore, our experiment compares \ourtool with the Original agent, both of which are exposed to the same degree of overfitting threat. Therefore, we believe this threat does not affect the validity of our findings.

%% file: sections/6_related.tex
\section{Related Work}
\label{sec:related_work}

\subsection{Prompt Reduction}

Previous approaches to prompt reduction improve the efficiency of LLMs by reducing the number of tokens in the prompt. The majority of prompt reduction work aims at Retrieval-Augmented Generation (RAG), where the system retrieves documents related to the user query from a database used as a reference by the LLM. FilCo~\cite{wang2023learning} filters the retrieved documents to retain only useful parts based on information-theoretic approaches. Selective Context~\cite{li2023compressing} calculates the self-information of each token from the output probability and removes tokens with lower self-information. RECOMP~\cite{xu2024recomp} introduces an abstractive compressor that paraphrases the compressed document into a concise abstract. Provence~\cite{chirkova2025provence} and LLMLingua-2~\cite{pan2024llmlingua2} train a classifier model to judge the relevance of each token in the document with ground truth generated by a teacher LLM. CPC~\cite{liskavets2025prompt} trains a context-aware sentence encoder to judge similarity based on the user query. These approaches target prompts in natural language, so they do not preserve the important structure in code and command outputs. For example, LLMLingua-2 does not preserve the full method names of the test cases in the example shown in \figref{motivation-casestudy}, leading to unrecognizable reduced content compared to the reflection module in \ourtool. The following text shows the compression result of LLMLingua-2 with the \code{rate} parameter set to 0.25, which controls its compression ratio to the same degree of the Keep\% of \ourtool.

\begin{infobox}{LLMLingua-2's compression result of the \code{pytest-dev\_\_pytest-6202} example}
\ldots{} \code{norecursedirs 33\%_testpaths 35\%_collect_file PASSED 36\%_directory 37\%_custom_repr_failure PASSED 39\%_collect_report_postprocessing 40\%_ignore_collect_path PASSED 41\%_ignore_collect_not_called 43\%_exclude_on_option 44\% 45\%_hooks PASSED 47\%_collect_file_dir 48\%_parsearg 50\%]_collect_topdir 51\%_collect_protocol_single_function 52\%_protocol_method 54\%_custom_nodes_multi_id 55\%_collect_subdir_event_ordering_commandline_args 58\%_serialization_byid 59\%_find_byid_instance_parents 60\%_global_file 62\%_pkgfile} \ldots{}
\end{infobox}

There are also approaches to prompt reduction specifically designed for code input. \citet{suneja2021probing} and \citet{rabin2022syntax} minimize the input code while preserving the output based on program simplification approaches. \citet{zhang2022diet} and \citet{wang2024natural} prune tokens in the code based on heuristic rules. \citet{yang2024less} compresses the docstring in the source code. \citet{pan2025hidden} reformats the code to remove tokens related to whitespace and indentation.

However, all approaches mentioned above aim to reduce prompts in the \emph{initial} input to the LLM, and do not consider the iterative nature of agent systems. Since trajectories are generated iteratively in agent systems, the timing of reduction becomes an important factor that can affect the efficiency and performance of the agent (as we show in \secref{impl-hyperparam}). Therefore, these approaches cannot be directly applied to agent trajectories without discussing the timing of prompt reduction.

\citet{lindenbauer2025complexity} propose an approach to trajectory reduction in parallel with our work. Their approach can be considered a variant of the Delete baseline (in \secref{impl-hyperparam-llm}) that deletes only the tool message but not the assistant message after a delay of 10 steps. Our results in \tabref{impl-hyperparam-result-llms} show that the Delete baseline is inferior to all LLM variants, with a severe decrease in Pass\% by 7\% and an increase in Step and PStep by 14\%. Furthermore, the choice of $a=10$ would cause significant overhead due to KV Cache invalidation.

\subsection{Context Management in Agent Systems}

Various agent systems employ some form of context management mechanism. For example, Cursor~\cite{cursor} and Claude Code~\cite{claude_code} compress the context with an LLM when the context window is full. Trae Agent~\cite{gao2025trae} truncates each tool response to the first 16 KB. SWE-agent~\cite{yang2024sweagent} has multiple configurable mechanisms such as removing text identified by a regular expression. These mechanisms are often ad-hoc and applied sparingly, aiming to improve the robustness of the agent in corner cases. In comparison, \ourtool aims to improve the overall efficiency of the agent by applying trajectory reduction aggressively. Such context management mechanisms are also not reported or evaluated as a first-class research question in existing papers~\cite{lindenbauer2025complexity}. In contrast, this paper provides an in-depth discussion and a large-scale evaluation of \ourtool.

\subsection{Efficiency Improvement for White-Box LLM Systems}

Some approaches improve the efficiency of LLMs by fine-tuning the model structure or optimizing the inference process. DMC~\cite{nawrot2024dynamic} combines adjacent KV Cache entries, which has the effect of merging multiple tokens into one token, thus saving computational cost. PISCO~\cite{louis2025pisco} distills RAG documents into embedding vectors with minimal cross-entropy to save tokens. Prune-on-Logic~\cite{zhao2025can} prunes the Chain-of-Thought tokens during model training to improve the reasoning efficiency of the trained model. MEM1~\cite{zhou2025mem1} trains an LLM that can modify the content in fixed-length memory. Alpine~\cite{saad2025adaptive} adds a pruning layer in each Transformer block to reduce computational cost.

Since these approaches modify the LLM's training or inference process, they require white-box access to the model. However, many state-of-the-art LLMs for agent systems are proprietary and do not permit such modifications. In comparison, \ourtool does not depend on the internal mechanisms of the LLM and can be applied to a wider range of agent systems.

%% file: sections/7_ending.tex
\section{Conclusion}

This paper addresses the efficiency concerns of coding LLM agents, an aspect often overlooked in existing studies. Through an analysis of trajectories in SWE-bench Verified, we identify typical sources of waste in these trajectories. We then propose \ourtool, an inference-time trajectory reduction approach designed to reduce the token cost of agent systems with a reflection module. We integrate \ourtool into a top-performing coding agent and investigate hyperparameter settings for the reflection module through quantitative experiments. Finally, we evaluate \ourtool on both SWE-bench Verified and Multi-SWE-bench Flash using Claude 4 Sonnet and Gemini 2.5 Pro. The results show that \ourtool consistently reduces input token usage by 39.9\%--59.7\%, resulting in significant computational cost savings of 21.1\%--35.9\%. Moreover, \ourtool maintains performance parity with the Original agent (ranging between -1.0\% and +2.0\%). This work demonstrates that the cost of LLM agents can be reduced without compromising performance using a simple approach. This approach benefits the practical application of agent systems and serves as a foundation for future research.

\section*{Data Availability}

The artifacts of this paper, including the implementation of \ourtool, scripts to render the tables and figures of the experimental results, and the raw trajectories collected in the experiments, are available at \href{https://doi.org/10.6084/m9.figshare.30073654}{https://doi.org/10.6084/m9.figshare.30073654}.

\begin{acks}
This work is sponsored by the National Key Research and Development Program of China under Grant No. 2022YFB4501902, the National Natural Science Foundation of China under Grant No. 92582202, and ByteDance Seed.
\end{acks}